\documentclass[preprint,12pt, numbers]{elsarticle}

\usepackage{amsmath,amssymb,amsfonts}
\usepackage{algorithmic}
\usepackage{graphicx}
\usepackage{subcaption}
\usepackage{textcomp}
\usepackage{xcolor}
\usepackage{natbib}
\setcitestyle{square, comma, number}
\usepackage{float}
\usepackage{color,soul}
\usepackage{etoolbox}
\usepackage{ulem}
\usepackage{tikz}
\usepackage[hyphens]{url}
\usepackage{array}
\usetikzlibrary{positioning, fit, arrows.meta, shapes}
\usepackage[latin1]{inputenc}
\usepackage{booktabs}
\usepackage{caption}
\usepackage{soul}

\usepackage{amssymb}

\usepackage{geometry}
\journal{arXiv}

\begin{document}

\begin{frontmatter}

\title{Dataset for Investigating Anomalies in Compute Clusters}

\author[1]{Diana McSpadden\corref{cor1}}
\ead{dianam@jlab.org}

\author[1]{Yasir Alanazi}
\ead{alanazi@jlab.org}

\author[1]{Bryan Hess}
\ead{bhess@jlab.org}

\author[1]{Laura Hild}
\ead{lsh@jlab.org}

\author[1]{Mark Jones}
\ead{maj@jlab.org}

\author[2]{Yiyang Lu}
\ead{ylu21@wm.edu}

\author[1]{Ahmed Mohammed}
\ead{ahmedm@jlab.org}

\author[1]{Wesley Moore}
\ead{wmoore@jlab.org}

\author[2]{Jie Ren}
\ead{jren03@wm.edu}

\author[1]{Malachi Schram}
\ead{schram@jlab.org}

\author[2]{Evgenia Smirni}
\ead{esmirni@cs.wm.edu}

\affiliation[1]{
            organization={Thomas Jefferson National Accelerator Laboratory},
            city={Newport News},
            state={VA 23606},
            country={USA}}

\affiliation[2]{organization={College of William and Mary},
            city={Williamsburg},
            state={VA 23187},
            country={USA}}

\begin{abstract}

The dataset was collected for 332 compute nodes throughout May 19 - 23, 2023. 
May 19 - 22 characterizes normal compute cluster behavior, while May 23 includes an anomalous event.
The dataset includes eight CPU, 11 disk, 47 memory, and 22 Slurm metrics.
It represents five distinct hardware configurations and contains over one million records, totaling more than 180GB of raw data.

\end{abstract}

\end{frontmatter}

\section{Background and Summary}

Motivated by the goal to develop a digital twin of a compute cluster, the dataset was collected using a Prometheus server \citep{prometheusoverview} scraping the Thomas Jefferson National Accelerator Facility (JLab) batch cluster used to run an assortment of physics analysis and simulation jobs, where analysis workloads leverage data generated from the laboratory's electron accelerator, and simulation workloads generate large amounts of flat data that is then carved to verify amplitudes.
Metrics were scraped from the cluster throughout May 19 - 23, 2023. 
Data from May 19 to May 22 primarily reflected normal system behavior, while May 23, 2023, recorded a notable anomaly. 
This anomaly was severe enough to necessitate intervention by JLab IT Operations staff.

The metrics were collected from CPU, disk, memory, and Slurm. 
Metrics related to CPU, disk, and memory provide insights into the status of individual compute nodes. 
Furthermore, Slurm metrics collected from the network have the capability to detect anomalies that may propagate to compute nodes executing the same job.

\section{Methods}

Agents on the hosts, \texttt{node\_exporter} and \texttt{prometheus-slurm-exporter}, collect the \texttt{node\_} (CPU, disk, and memory) and \texttt{slurm\_} metrics, respectively.  The \texttt{node\_exporter} generally pulls its information from Linux's \texttt{proc} and \texttt{sysfs}, and \texttt{prometheus-slurm-exporter} from Slurm's command-line utilities, but specifics can be found in their (open) source code.\citep{nodeexp,slurmexp}

Prometheus scrapes the metrics from the exporters over the network and stores them in a time-series database.  Scrapes target every thirty seconds for Slurm metrics and once a minute for other metrics, but samples may be missing if hosts or their exporters are down, otherwise unable to promptly respond to network requests (e.g. because of a saturated uplink), or unable to promptly collect metrics (e.g. if the Slurm controller is so busy that it is not responsive to commands).

A Python program using the \texttt{prometheus-api-client} \citep{pythonpromapiclient} package queries Prometheus and exports a selection of the information. 
There are two major methods of the prometheus client object  used to fetch data from the prometheus server: 
\begin{itemize}
    \item \texttt{get\_metric\_range\_data}
    \item and \texttt{custom\_query\_range}.
\end{itemize}
Both methods, when used with the \texttt{MetricRangeDataFrame} function, output a pandas\\ DataFrame with the timestamp set as index. 

The \texttt{get\_metric\_range\_data} method outputs the raw readings and timestamps of the provided metric in the interval [\texttt{start\_time}, \texttt{end\_time}]. 
The \texttt{custom\_query\_range} method is more customizable. It evaluates the provided query at all evaluation timestamps in the interval [\texttt{start\_time} : \texttt{end\_time} : \texttt{step}]. 
At each evaluation timestamp, it has a default lookback of 5 minutes. Over this lookback window, the most recent reading is assigned the evaluation timestamp. Assigning all readings to a a certain timestamp is a useful property when time allignment is required across different variables or compute nodes. To get the cpu rates directly over a defined lookback interval (144 seconds) for specific nodes (all farm nodes in addition to nodes A, and B), we use the following custom query:
\\
\newline \texttt \textbraceleft rate(node\_cpu\_seconds\_total\textbraceleft instance=$\sim$"farm.*|A|B"\ \textbraceright[144s])\textbraceright .

\section{Data Records}
 The dataset contains directories for each of the metric types. These are described in the following sections \ref{sec:cpu_data} - \ref{sec:slurm_data}.

\subsection{CPU data} \label{sec:cpu_data}
The CPU data files are sorted by the datetime of the recorded metric. For example,  the file `node\_cpu\_seconds\_total\_0.csv' contains the first recorded observations for an event followed by `node\_cpu\_seconds\_total\_1.csv'.

\begin{table}
\centering
\scriptsize
\begin{tabular}{llllllll}
\hline

  & \_\_name\_\_            & cpu        & instance   & job       & mode  & timestamp                     & value \\ \hline
0 & node\_cpu\_seconds\_total & 0 & farm140105:9100 & node & idle & 2023-05-19 04:00:52.737999872 & 102923.11  \\
1 & node\_cpu\_seconds\_total & 0 & farm140105:9100 & node & iowait & 2023-05-19 04:00:52.737999872 & 781.04  \\
2 & node\_cpu\_seconds\_total & 0 & farm140105:9100 & node & irq & 2023-05-19 04:00:52.737999872 & 0.0 

\\ \hline
\end{tabular}
\caption{Header and first three rows of node\_cpu\_seconds\_total\_0.csv from the May event.}
\label{tab:cpu_data}
\end{table}

Table \ref{tab:cpu_data} illustrates the header and first three rows of the first CPU data file from the May event.
The value identifies the amount of time the CPU has spent performing the specified task mode. The numbers reported are counters/aggregates since the start of this year or the time when the system booted in this year. 

\begin{itemize}
    \item An unlabeled index column.
    \item \textbf{\_\_name\_\_}: The description for the collected data.
    \item \textbf{cpu} The CPU index for the indicated \textbf{instance}.
    \item \textbf{instance}: The node instance on the compute cluster.
    \item \textbf{job}: ``node".
    \item \textbf{mode}: a type of task being performed by the CPU:
        \begin{itemize}
            \item \textbf{user}: Time spent with normal processing in user mode.
            \item \textbf{nice}: Time spent with niced processes in user mode.
            \item \textbf{system}: Time spent running in kernel mode.
            \item \textbf{idle}: Time spent in vacations twiddling thumbs.
            \item \textbf{iowait}: 	Time spent waiting for I/O to completed. This is considered idle time too.
            \item \textbf{irq}: Time spent serving hardware interrupts. See the description of the intr line for more details.
            \item \textbf{softirq}: Time spent serving software interrupts.
            \item \textbf{steal}: Time stolen by other operating systems running in a virtual environment.
            \item \textbf{guest}: Time spent for running a virtual CPU or guest OS under the control of the kernel.
        \end{itemize}
    \item \textbf{timestamp}: the time that Prometheus collected the value.
    \item \textbf{value}: Amount of time the CPU has spent performing the specified task mode.
\end{itemize}

\subsection{Disk data} \label{sec:disk_data}

The disk data files are sorted by disk metric name, and the datetime of the recorded metric, i.e. the file `node\_disk\_read\_bytes\_total\_0.csv' contains the first recorded disk\_read\_bytes observations for an event followed by `node\_disk\_read\_bytes\_total\_1.csv', etc.

The following disk metrics are included in the dataset in separate files.

\begin{itemize}
    \item \textbf{node\_disk\_io\_now}: (field 9) The only field that should go to zero. Incremented as requests are given to appropriate struct request\_queue and decremented as they finish. 
    \\Example file name: node\_disk\_io\_now\_0.csv
    
    \item \textbf{node\_disk\_io\_time\_seconds\_total}: (field 10) Number of seconds spent doing I/Os. This field increases so long as field 9 is nonzero. 
    \\Example file name: node\_disk\_io\_time\_seconds\_total\_0.csv
    
    \item \textbf{node\_disk\_io\_time\_weighted\_seconds\_total}: (field 11) Weighted \# of seconds spent doing I/Os.
    This field is incremented at each I/O start, I/O completion, I/O merge, or read of these stats by the number of I/Os in progress (field 9) times the number of seconds spent doing I/O since the last update of this field.  This can provide an easy measure of both I/O completion time and the backlog that may be accumulating. 
    \\Example file name: node\_disk\_io\_time\_weighted\_seconds\_total\_0.csv
    
    \item \textbf{node\_disk\_read\_bytes\_total}: (field 3) This is the total number of bytes read successfully. \\Example file name: node\_disk\_read\_bytes\_total\_0.csv
    
    \item \textbf{node\_disk\_reads\_completed\_total}: (field 1) This is the total number of reads completed successfully.\\ Example file name: node\_disk\_reads\_completed\_total\_0.csv

    \item \textbf{node\_disk\_reads\_merged\_total}: (field 2) \# of merged. Reads which are adjacent to each other may be merged for efficiency.  Thus two 4K reads may become one 8K read before it is ultimately handed to the disk, and so it will be counted (and queued) as only one I/O.  This field lets you know how often this was done.
    \\ Example file name: node\_disk\_reads\_merged\_total\_0.csv

    \item \textbf{node\_disk\_read\_time\_seconds\_total}: (field 4) This is the total number of seconds spent by all reads (as measured from \_\_make\_request() to end\_that\_request\_last()).
    \\ Example file name: node\_disk\_read\_time\_seconds\_total\_0.csv

    \item \textbf{node\_disk\_writes\_completed\_total}: (field 5) This is the total number of writes completed successfully.
    \\ Example file name: node\_disk\_writes\_completed\_total\_0.csv

    \item \textbf{node\_disk\_writes\_merged\_total}: (field 6) \# of merged. Writes which are adjacent to each other may be merged for efficiency. This field lets you know how often this was done.
    \\ Example file name: node\_disk\_writes\_merged\_total\_0.csv

    \item \textbf{node\_disk\_write\_time\_seconds\_total}: (field 8) This is the total number of seconds spent by all writes (as measured from \_\_make\_request() to end\_that\_request\_last()).
    \\ Example file name: node\_disk\_write\_time\_seconds\_total\_0.csv

    \item \textbf{node\_disk\_written\_bytes\_total}: (field 7) This is the total number of bytes written successfully.
    \\ Example file name: node\_disk\_written\_bytes\_total\_0.csv
\end{itemize}

\begin{table}
\centering
\scriptsize
\begin{tabular}{llllllll}
\hline

  & \_\_name\_\_           & device        & instance   & job        & timestamp                     & value \\ \hline
3808 & node\_disk\_io\_now & md126 & farm140105:9100 & node & 2023-05-19 04:00:52.737999872 & 0.0  \\
3809 & node\_disk\_io\_now & md126 & farm140108:9100 & node & 2023-05-19 04:00:50.910000128 & 0.0  \\
3810 & node\_disk\_io\_now & md126 & farm140109:9100 & node & 2023-05-19 04:00:52.364999936 & 0.0  
\\ \hline
\end{tabular}
\caption{Header and first three rows of node\_disk\_io\_now\_0.csv from the May event.}
\label{tab:disk_data}
\end{table}

Refer to field numbers in Kernel Documentation \citep{iostatsdoc} for further information.

\subsection{Memory data} \label{sec:memory_data}

Memory metrics are stored in separate files, one for each metric identified by [METRIC\_NAME].csv. Values recorded were scraped by Prometheus at the time recorded in the CSV, and may increase or decrease. For filenames ending in \texttt{\_bytes}, while the underlying \texttt{meminfo} values are measured in kibibytes, \texttt{node\_exporter} has multiplied them out in order to suit the Prometheus convention of using base units, so they are indeed bytes.

The following memory metrics are provided in the dataset:
\begin{itemize}
    \item \textbf{node\_memory\_Active\_anon\_bytes}: 
    Subset of Active which is \\ ``anonymous," i.e. not backed by a filesystem.
    
    \item \textbf{node\_memory\_Active\_bytes}: 
    Memory that has been used more recently and usually not reclaimed unless absolutely necessary.
    
    \item \textbf{node\_memory\_Active\_file\_bytes}: 
    Subset of Active which is backed by a filesystem.
    
    \item \textbf{node\_memory\_AnonHugePages\_bytes}: 
    Non-file backed huge \\ pages mapped into user-space page tables.
    
    \item \textbf{node\_memory\_AnonPages\_bytes}:
    Non-file backed pages mapped into user-space\\ pagetables.
    
    \item \textbf{node\_memory\_Buffers\_bytes}:
    Relatively temporary storage for raw disk blocks that shouldn't get tremendously large (20 MB or so).
    
    \item \textbf{node\_memory\_Cached\_bytes}:
    In-memory cache for files read from the disk (the page cache).  Doesn't include SwapCached.
    
    \item \textbf{node\_memory\_CmaFree\_bytes}:
    Free CMA (Contiguous Memory Allocator) pages (bytes).
    
    \item \textbf{node\_memory\_CmaTotal\_bytes}:
    Total CMA (Contiguous Memory Allocator) pages (bytes).
    
    \item \textbf{node\_memory\_CommitLimit\_bytes}:
    This is the total amount of memory currently available to be allocated on the system.
    
    \item \textbf{node\_memory\_Committed\_AS\_bytes}:
    The amount of memory presently allocated on the system.  The committed memory is a sum of all of the memory which has been allocated by processes, even if it has not been ``used" by them as of yet.
    
    \item \textbf{node\_memory\_DirectMap1G\_bytes}:
    Breakdown of page table sizes (1G) used in the kernel's identity mapping of RAM.
    
    \item \textbf{node\_memory\_DirectMap2M\_bytes}:
    Breakdown of page table sizes (2M) used in the kernel's identity mapping of RAM.
    
    \item \textbf{node\_memory\_DirectMap4k\_bytes}:
    Breakdown of page table sizes (4k) used in the kernel's identity mapping of RAM.
    
    \item \textbf{node\_memory\_Dirty\_bytes}:
    Memory which is waiting to get written back to the disk.
    
    \item \textbf{node\_memory\_HardwareCorrupted\_bytes}:
    The amount of RAM/memory in KB, the kernel identifies as corrupted.
    
    \item \textbf{node\_memory\_HugePages\_Free}:
    The number of huge pages in the pool that are not yet allocated.
    
    \item \textbf{node\_memory\_HugePages\_Rsvp}:
    Short for "reserved," this is the number of huge pages for which a commitment to allocate from the pool has been made, but no allocation has yet been made. Reserved huge pages guarantee that an application will be able to allocate a huge page from the pool of huge pages at fault time.
    
    \item \textbf{node\_memory\_HugePages\_Surp}:
    Short for ``surplus," this is the number of huge pages in the pool above the value in \texttt{/proc/sys/vm/nr\_hugepages}.
    
    \item \textbf{node\_memory\_HugePages\_total}:
    The size of the pool of huge pages.
    
    \item \textbf{node\_memory\_Hugepagesize\_bytes}:
    The default hugepage size.
    
    \item \textbf{node\_memory\_Inactive\_anon\_bytes}: 
    Subset of Inactive which is ``anonymous," i.e. not backed by a filesystem.
    
    \item \textbf{node\_memory\_Inactive\_bytes}:
    Memory which has been less recently used. It is more eligible to be reclaimed for other purposes.
    
    \item \textbf{node\_memory\_Inactive\_file\_bytes}: 
    Subset of Inactive which is backed by a filesystem.
    
    \item \textbf{node\_memory\_KernelStack\_bytes}:
    Memory consumed by the kernel stacks of all tasks.
    
    \item \textbf{node\_memory\_Mapped\_bytes}:
    Files which have been mapped, such as libraries.
    
    \item \textbf{node\_memory\_MemAvailable\_bytes}:
    An estimate of how much memory is available for starting new applications, without swapping. Calculated from MemFree, SReclaimable, the size of the file LRU lists, and the low watermarks in each zone. The estimate takes into account that the system needs some page cache to function well, and that not all reclaimable slab will be reclaimable, due to items being in use. The impact of those factors will vary from system to system.
    
    \item \textbf{node\_memory\_MemFree\_bytes}:
    Total free RAM. On highmem systems, the sum of LowFree+HighFree.
    
    \item \textbf{node\_memory\_MemTotal\_bytes}:
    Total usable RAM (i.e. physical RAM minus a few reserved bits and the kernel binary code).
    
    \item \textbf{node\_memory\_Mlocked\_bytes}:
    Memory locked with mlock().
    
    \item \textbf{node\_memory\_NFS\_Unstable\_bytes}:
    Pages which have been written to an NFS or Lustre server, but have not been committed to stable storage.
    
    \item \textbf{node\_memory\_PageTables\_bytes}:
    Memory consumed by userspace page tables.
    
    \item \textbf{node\_memory\_Percpu\_bytes};
    Memory allocated to the percpu allocator used to back percpu allocations. This stat excludes the cost of metadata.
    
    \item \textbf{node\_memory\_Shmem\_bytes}:
    Total memory used by shared memory (shmem) and tmpfs.
    
    \item \textbf{node\_memory\_Slab\_bytes}:
    In-kernel data structures cache
    
    \item \textbf{node\_memory\_SReclaimable\_bytes}:
    Part of Slab, that might be reclaimed, such as caches.
    
    \item \textbf{node\_memory\_SUnreclaim\_bytes}:
    Part of Slab, that cannot be reclaimed on memory pressure.
    
    \item \textbf{node\_memory\_SwapCached\_bytes}:
    Memory that once was swapped out, is swapped back in but still also is in the swapfile (if memory is needed it doesn't need to be swapped out AGAIN because it is already in the swapfile. This saves I/O).
    
    \item \textbf{node\_memory\_SwapFree\_bytes}:
    Memory which has been evicted from RAM, and is temporarily on the disk.
    
    \item \textbf{node\_memory\_SwapTotal\_bytes}:
    Total amount of swap space available.
    
    \item \textbf{node\_memory\_Unevictable\_bytes}:
    Memory allocated for userspace which cannot be reclaimed, such as mlocked pages, ramfs backing pages, secret memfd pages etc.
    
    \item \textbf{node\_memory\_VmallocChunck\_bytes}:
    Largest contiguous block of vmalloc areas\\ which is free.
    
    \item \textbf{node\_memory\_VmallocTotal\_bytes}:
    Total size of vmalloc virtual address space.
    
    \item \textbf{node\_memory\_VmallocUsed\_bytes}:
    Amount of vmalloc area which is used.
    
    \item \textbf{node\_memory\_Writeback\_bytes}:
    Memory which is actively being written back to the disk.
    
    \item \textbf{node\_memory\_WritebackTmp\_bytes}:
    Memory used by FUSE for temporary writeback buffers.
\end{itemize}

\begin{table}
\centering
\scriptsize
\begin{tabular}{llllllll}
\hline

  & \_\_name\_\_            & instance        & job   & timestamp                     & value \\ \hline
11 & node\_memory\_Active\_bytes & farm140105:9100 & node & 2023-05-19 04:00:52.737999872 & 11222962176.0 \\
12 & node\_memory\_Active\_bytes & farm140108:9100 & node & 2023-05-19 04:00:50.910000128 & 19978526720.0 \\
13 & node\_memory\_Active\_bytes & farm140109:9100 & node & 2023-05-19 04:00:52.364999936 & 19792236544.0
\\ \hline
\end{tabular}
\caption{Header and first three rows of node\_memory\_Active\_bytes.csv from the May event.}
\label{tab:memory_data}
\end{table}

Table \ref{tab:memory_data} illustrates the header and first three rows of the node\_memory\_Active\_bytes.csv. The 47 memory files included in the dataset use the same header and file structure:

\begin{itemize}
    \item An unlabeled index column.
    \item \textbf{\_\_name\_\_}: The filename and the description for the collected metric.
    \item \textbf{instance}: The name of the node on the compute cluster.
    \item \textbf{job}: ``node".
    \item \textbf{timestamp}: The time that Prometheus collected the value.
    \item \textbf{value}: The value of the \_\_name\_\_ metric for the instance at the timestamp.
\end{itemize}

\subsection{Slurm data} \label{sec:slurm_data}

The Slurm data files are grouped by the state of the CPU: Allocated, Idle, Other and Total.
\begin{itemize}
    \item \textbf{slurm\_node\_cpu\_alloc.csv}: Allocated CPUs, i.e. CPUs which have been allocated to a job.
    \item \textbf{slurm\_node\_cpu\_idle.csv}: Idle CPUs, i.e. CPUs not allocated to a job and thus available for use.
    \item \textbf{slurm\_node\_cpu\_other.csv}: Other CPUs, i.e. CPUs which are unavailable for use at the moment.
    \item \textbf{slurm\_node\_cpu\_total.csv}: Total CPUS, i.e. the total number of CPUs.
\end{itemize}

\begin{table}
\centering
\scriptsize
\begin{tabular}{llllllll}
\hline

  & \_\_name\_\_            & instance        & job   & node       & status  & timestamp                     & value \\ \hline
0 & slurm\_node\_cpu\_alloc & enpslurm21:9341 & slurm & farm140105 & mixed & 2023-05-19 04:00:47.217999872 & 42.0  \\
1 & slurm\_node\_cpu\_alloc & enpslurm21:9341 & slurm & farm140108 & mixed & 2023-05-19 04:00:47.217999872 & 36.0  \\
2 & slurm\_node\_cpu\_alloc & enpslurm21:9341 & slurm & farm140109 & mixed & 2023-05-19 04:00:47.217999872 & 34.0  
\\ \hline
\end{tabular}
\caption{Header and first three rows of slurm\_node\_cpu\_alloc.csv from the May event.}
\label{tab:slurm_data}
\end{table}

Table \ref{tab:slurm_data} illustrates the header and first three rows of an \textit{Allocated} slurm data file. The \textit{Idle}, \textit{Other}, and \textit{Other} CSV files use the same header and structure:

\begin{itemize}
    \item An unlabeled index column.
    \item \textbf{\_\_name\_\_}: The description for the collected data.
    \item \textbf{instance}: The hostname of the monitoring service.
    \item \textbf{job}: ``slurm".
    \item \textbf{node}: the name of the node on the compute cluster.
    \item \textbf{status}: the state of the node for which the \textbf{value} column count applies at the \textbf{timestamp}. Status can include the following suffixes and states:
        \begin{itemize}
            \item Suffixes:
                \begin{itemize}
                    \item \textbf{*}: The node is presently not responding and will not be allocated any new work. If the node remains non-responsive, it will be placed in the DOWN state (except in the case of \textbf{completing}, \textbf{drained}, \textbf{draining}, \textbf{fail}, \textbf{failing} nodes).
                    \item \textbf{\~}: The node is presently in powered off.
                    \item \textbf{\#}: The node is presently being powered up or configured.
                    \item \textbf{!}: The node is pending power down.
                    \item \textbf{\%}: The node is presently being powered down.
                    \item \textbf{\$}: The node is currently in a reservation with a flag value of ``maintenance.''
                    \item \textbf{@}: The node is pending reboot
                    \item \textbf{\^}: The node reboot was issued.
                    \item \textbf{\-}: The node is planned by the backfill scheduler for a higher priority job.
                    
                \end{itemize}
            \item States:
                \begin{itemize}
                    \item \textbf{allocated}: nodes which has been allocated to one or more jobs.
                    \item \textbf{allocated+}: The node is allocated to one or more active jobs plus one or more jobs are in the process of completing.
                    \item \textbf{completing}: All jobs associated with this node are in the process of completing. This node state will be removed when all of the job's processes have terminated and the Slurm epilog program (if any) has terminated.
                    \item \textbf{down}: The node is unavailable for use. Slurm can automatically place nodes in this state if some failure occurs. System administrators may also explicitly place nodes in this state. If a node resumes normal operation, Slurm can automatically return it to service.
                    \item \textbf{drained}: The node is unavailable for use per system administrator request.
                    \item \textbf{draining}: The node is currently executing a job, but will not be allocated additional jobs. The node state will be changed to state drained when the last job on it completes. Nodes enter this state per system administrator request.
                    \item \textbf{fail}: these nodes are expected to fail soon and are unavailable for use per system administrator request.
                    \item \textbf{failing}: The node is currently executing a job, but is expected to fail soon and is unavailable for use per system administrator request.
                    \item \textbf{future}: The node is currently not fully configured, but expected to be available at some point in the indefinite future for use.
                    \item \textbf{idle}: nodes not allocated to any jobs and thus available for use.
                    \item \textbf{inval}: The node did not register correctly with the controller. This happens when a node registers with less resources than configured in the slurm.conf file. The node will clear from this state with a valid registration (i.e. a slurmd restart is required).
                    \item \textbf{maint}: nodes which are currently marked with the maintenance flag.
                    \item \textbf{reboot\_issued}: A request to reboot this node has been made, but hasn't been handled yet.
                    \item \textbf{mixed}: nodes that have some of their CPUs allocated while others are idle.
                    \item \textbf{perfctrs (npc)}: Network Performance Counters associated with this node are in use, rendering this node as not usable for any other jobs.
                    \item \textbf{planned}: The node is planned by the backfill scheduler for a higher priority job.
                    \item \textbf{power\_down}: The node is pending power down.
                    \item \textbf{powered\_down}: The node is currently powered down and not capable of running any jobs.
                    \item \textbf{powering\_down}: The node is in the process of powering down and not capable of running any jobs.
                    \item \textbf{powering\_up}: The node is in the process of being powered up.
                    \item \textbf{reserved}: The node is in an advanced reservation and not generally available.
                    \item \textbf{unknown}: The Slurm controller has just started and the node's state has not yet been determined.
                \end{itemize}
        \end{itemize}
    \item \textbf{timestamp}: the time that Prometheus collected the value.
    \item \textbf{value}: Number of CPUs in the file state with the indicated status on the node at the timestamp.
    
\end{itemize}

\subsection{Encoding of the Compute Node Hardware Configuration}

Five distinct hardware configurations are represented in the dataset. The primary distinction is the number of CPUs on the node. The hardware configuration is encoded in the first six characters of the `instance' or `node' column in a CSV file. The five possible values are: [`farm14', `farm16', `farm18', `farm19', `farm23'].

\section{Usage Notes}

While the data from May 19 - 22 characterizes normal compute cluster behavior, and May 23 includes anomalous observations, the dataset \textbf{cannot be considered labeled data}. 
The set of nodes and the exact start and end time affected nodes demonstrate abnormal effects are unclear.
Thus, the dataset could be used to develop unsupervised machine-learning algorithms to detect anomalous events in a batch cluster.

\section{Acknowledgements}
This material is based upon work supported by the U.S. Department of Energy, Office of Science, Office of Nuclear Physics under contract DE-AC05-06OR23177.

 \bibliographystyle{plainnat} 
 \bibliography{main}

\end{document}